\documentclass[fleqn,usenatbib,letters]{mnras}

\usepackage{newtxtext,newtxmath}
\usepackage[T1]{fontenc}

\DeclareRobustCommand{\VAN}[3]{#2}
\let\VANthebibliography\thebibliography
\def\thebibliography{\DeclareRobustCommand{\VAN}[3]{##3}\VANthebibliography}

\usepackage{graphicx}	% Including figure files
\usepackage{amsmath}	% Advanced maths commands

\title[An extremely thin gravitational arc]{An extended and extremely thin gravitational arc from a lensed compact symmetric object at redshift 2.059}

\author[J. P. McKean et al.]{
J. P. McKean,$^{1,2,3}$\thanks{E-mail: mckean@astro.rug.nl}
C. Spingola,$^{4}$
D. M. Powell$^{5}$ 
and S. Vegetti$^{5}$
\\
$^{1}$Kapteyn Astronomical Institute, University of Groningen, Postbus 800, NL-9700 AV Groningen, The Netherlands\\
$^{2}$South African Radio Astronomy Observatory (SARAO), P.O. Box 443, Krugersdorp 1740, South Africa\\
$^{3}$Department of Physics, University of Pretoria, Lynnwood Road, Hatfield, Pretoria, 0083, South Africa\\
$^{4}$INAF--Istituto di Radioastronomia, Via Gobetti 101, I-40129, Bologna, Italy\\
$^{5}$Max Planck Institute for Astrophysics, Karl-Schwarzschild-Stra\ss{}e 1, 85748 Garching bei M\"unchen, Germany}

\date{Accepted 2025 April 14. Received 2025 April 13; in original form 2025 January 20}

\pubyear{\the\year{}}

\begin{document}
\label{firstpage}
\pagerange{\pageref{firstpage}--\pageref{lastpage}}
\maketitle

% Abstract of the paper
\begin{abstract}
Compact symmetric objects (CSOs) are thought to be short-lived radio sources with two lobes of emission that are separated by less than a kpc in projection. However, studies of such systems at high redshift is challenging due to the limited resolution of present-day telescopes, and can be biased to the most luminous objects. Here we report imaging of a gravitationally lensed CSO at a redshift of 2.059 using very long baseline interferometry at 1.7 GHz. The data are imaged using Bayesian forward modelling deconvolution, which reveals a spectacularly extended and thin gravitational arc, and several resolved features within the lensed images. The surface brightness of the lensing-corrected source shows two mini-lobes separated by 642~pc in projection, with evidence of multiple hotspots that have brightness temperatures of $10^{8.6}$ to $10^{9.2}$~K, and a total luminosity density of $10^{26.3}$~W~Hz$^{-1}$. By combining the well-resolved radio source morphology with previous multi-wavelength studies, we conclude that this object is likely a CSO of type 2, and that the properties are consistent with the bow-shock model for compact radio sources. Our analysis highlights the importance of combining high quality data sets with sophisticated imaging and modelling algorithms for studying the high redshift Universe.
\end{abstract}

% Select between one and six entries from the list of approved keywords.
% Don't make up new ones.

%%%%%%%%%%%%%%%%%%%%%%%%%%%%%%%%%%%%%%%%%%%%%%%%%%

%%%%%%%%%%%%%%%%% BODY OF PAPER %%%%%%%%%%%%%%%%%%

\section{Introduction}
\label{sec:intro}

Active Galactic Nuclei (AGN) with prominent radio jets are thought to have a profound effect on the formation and evolution of their host galaxies, through regulating the amount of cold molecular gas available for star formation (e.g. \citealt{kondapally2023}). In addition, the host galaxy and/or wide precession angles can also frustrate the expansion of the radio jets as they propagate through the dense interstellar medium (ISM; e.g. \citealt{stanghellini2024}). Therefore, detailed studies of powerful radio sources at cosmologically important epochs, such as when black hole growth and the star-formation density of the Universe were highest, is needed for testing models for the evolution of radio-mode AGN activity. Also, at these epochs, galaxies are both intrinsically smaller and difficult to resolve, which makes imaging their surface brightness distributions challenging (e.g. \citealt{muxlow2020,sweijen2022}). It is for this reason that many detailed studies of the high redshift Universe rely on gravitational lensing, often through using a simple estimate of the magnification, to infer the properties of the background object. However, with the recent advancement of sophisticated lens modelling techniques (e.g. \citealt{powell2021}) and high-quality data sets taken at high angular resolution (e.g. \citealt{spingola2018,stacey2024}), we are now in a position to make robust images of the high redshift Universe, after correcting for the distortion of the gravitational lensing (e.g. \citealt{stacey2025}). 

In this letter, we present high angular resolution imaging of the gravitational lens system JVAS B1938+666, using very long baseline interferometry (VLBI), which shows the most distinct gravitational arc ever observed. JVAS B1938+666 was discovered by \citet{king1997} as part of the Jodrell Bank--Very Large Array Astrometric Survey (JVAS; \citealt{king1999}). This particular system has been well-studied to date because it comprises a foreground massive elliptical galaxy at redshift $z_l = 0.8809\pm0.0005$ \citep{tonry2000} that gravitationally lenses a background object at redshift $z_s = 2.0590\pm0.0003$ \citep{riechers2011,spingola2020} into an almost complete Einstein ring at near-infrared wavelengths \citep{king1998,lagattuta2012}. At radio wavelengths, the source has two components, where one is doubly imaged and the other forms four images, with an extended gravitational arc connecting three of those lensed images \citep{king1997,spingola2020}. The large number of observational constraints for this system has allowed detailed mass modelling of the lens \citep{lagattuta2012}, and the discovery of a low mass halo along the line-of-sight to the background source \citep{vegetti2012}.

Here, the visibility data from new global VLBI observations of JVAS B1938+666 are imaged using a novel Bayesian forward modelling deconvolution process, which also provides a high angular resolution model for the surface brightness distribution of the un-lensed radio source. In Section \ref{sec:obs}, we present our observations, data reduction and modelling procedure. The analysis of the image-plane deconvolved imaging and the source-plane resolved radio source is presented in Section \ref{sec:results}. Finally, we discuss our results and present our conclusions in Sections \ref{sec:disc} and \ref{sec:conc}, respectively. 

Throughout, we assume a flat $\Lambda$CDM cosmology \citep{Planck2020}.

\section{Observations \& Modelling}
\label{sec:obs}

In this section, we present new VLBI observations of JVAS~B1938+666, and a brief overview of the method used to image the visibility data and produce a resolved model for the background radio source.

\begin{figure*}
    \centering
    \includegraphics[height=0.64\textheight]{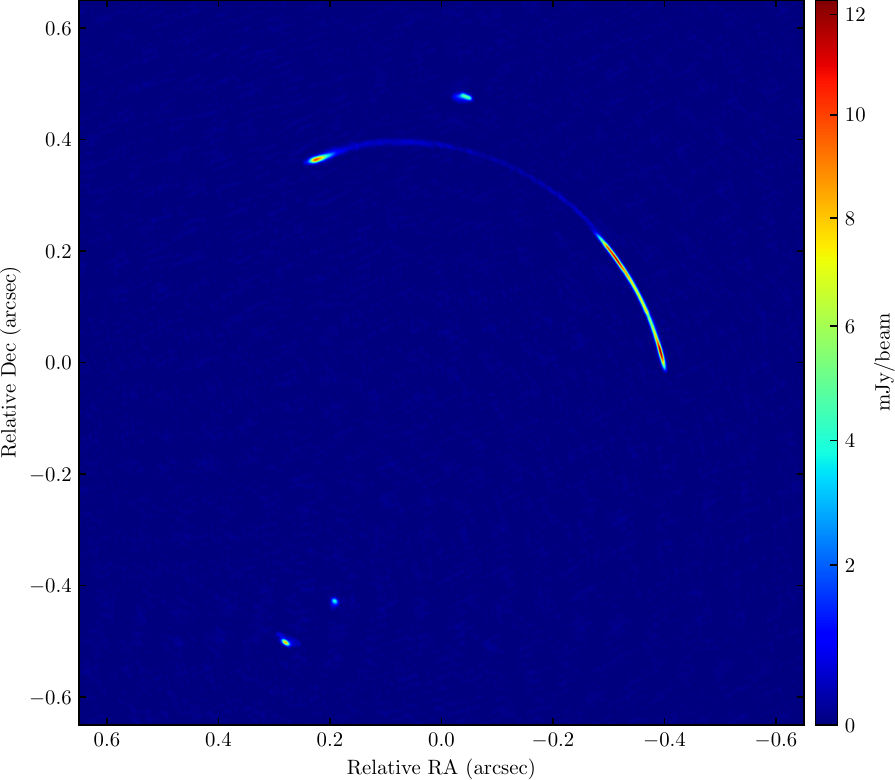}
    \caption{
    The Bayesian forward modelling deconvolution of the global VLBI imaging of JVAS B1938+666 at 1.7 GHz. The image has an rms of 34.3~$\mu$Jy~beam$^{-1}$ and the deconvolved image has been made with a restoring beam of $7.4\times4.7$~mas$^{2}$ at a position angle of 32.1~deg east of north. The centre of the image is ${\rm RA}=19^{\rm h}38^{\rm m}25\fs3407$, ${\rm Dec}= +66\degr48\arcmin52\farcs809\,40$ (J2000).}
    \label{fig:1938model}
\end{figure*}

\subsection{Global VLBI observations of JVAS B1938+666}

JVAS B1938+666 was observed at 1.7 GHz with the global VLBI array on 2011 November 6 for a total of 14 h (GM068; PI: McKean). The array comprised 11 stations from the European VLBI Network (EVN), the 10 stations of the Very Long Baseline Array (VLBA) and the Green Bank Telescope. However, due to logistical problems, the data from the VLBA stations at North Liberty, Ford Davis and Kitt Peak were completely lost, and the data from Owens Valley, Brewster and Mauna Kea were partially lost. The part of the observations that only used the VLBA were phase referenced with the nearby calibrator J1933+654, and the bright calibrator sources 3C454.3 and 3C345 were observed every $\sim4$~h to act as fringe finders during the data correlation process and for determining the bandpass during the data reduction stage. The recording rate was 512 Mbits\,s$^{-1}$, which produced 8 spectral windows with 8 MHz bandwidth each and 2 polarizations (RR and LL). The data from the 19 stations were correlated at the Joint Institute for VLBI--European Research Infrastructure Consortium (JIV--ERIC), where each 8 MHz spectral window was divided into 32 spectral channels and a visibility averaging time of 2 s was applied. 

The processing of the data was carried out within {\sc aips} (Astronomical Imaging Processing Software; \citealt{greisen2003}) using standard methods for VLBI observations. In particular, the calibrator source J1933+654 was used to derive the rates and delays for the phase referenced portion of the data that was taken with the VLBA. From this, an initial model for JVAS~B1938+666 was generated to determine the rates and delays for the entire data set. This process was then iteratively performed until no major improvement in the sky model was found. This model was then used as the starting point for an iterative process of amplitude and phase self-calibration within {\sc DIFMAP} (Difference Mapping Package; \citealt{shepherd1997}) using the central 2-MHz portion of each spectral window, to avoid bandwidth and time smearing. The phase self-calibration had a minimum solution interval of 30~s, whereas the amplitude self-calibration used a solution interval of 60 min. For amplitude self-calibration solution intervals shorter than this, the results became unstable because of the limited number of antennas available at any given time. The solutions were applied to all channels within each spectral window. As a final step, the data weights were re-calculated using the rms scatter for each baseline over a 5 min tine interval.

\subsection{Modelling procedure}
The model for the surface brightness emission in the image plane and in the un-lensed source plane is determined using {\sc pronto} \citep{vegetti2009, rybak2015a, rizzo2018, ritondale2019, powell2021, powell2022, ndiritu2024}, which fits directly to the visibilities within a hierarchical Bayesian framework.

In this framework, the source surface brightness distribution, $\bmath{s}$, is gravitationally lensed forward using a lensing operator, $\bmath{\mathsf{L}}$, such that the model image-plane surface brightness distribution is given by $\bmath{\mathsf{L}}(\bmath{\eta})\,\bmath{s}$. The parameters of the lensing operator, $\bmath{\eta}$, describe the mass model of the lens, which maps a position in the source plane to the multiple images that are magnified and distorted in the image plane. Here, the lensing operator can be thought of as the optics of the system. Finally, the data are observed through the interferometer, which results in a filtering of the structure due to the incomplete sampling of the Fourier plane. This is given by $\bmath{\mathsf{D}}$, and its Fourier transform is the point spread function (psf) of the interferometer. Within $\bmath{\mathsf{D}}$ is the sampling and weighting of the data, for example, via a $uv$-taper or some weighting-scheme based on the noise properties of the visibilities. Throughout, we apply a natural weighting of the visibilities. The model of the data, $\bmath{d}$, is therefore described by,
\begin{equation}
    \bmath{d} = \bmath{\mathsf{D}}\, \bmath{\mathsf{L}}(\bmath{\eta})\, \bmath{s} + \bmath{n},
\label{eq:datamodel}
\end{equation}
where $\bmath{n}$ is the visibility noise, which is assumed to be Gaussian and un-correlated, and has a covariance given by $\bmath{\mathsf{C}}^{-1}$. An extensive discussion on the model used for the analysis of interferometric data within this Bayesian hierarchical framework is given by \citet{powell2021}.

A key input to equation (\ref{eq:datamodel}) is the parametrization of the source surface brightness distribution. Within the traditional {\sc clean} algorithm, and its derivatives, there is a parametric description of the source, which is typically a collection of delta functions or truncated Gaussian functions. These simple models ensure a smoothness, but have limited flexibility to describe sources with complex morphologies. Here, we use a pixelated model for the source surface brightness distribution, which has the advantage of having sufficient flexibility, but requires additional constraints or priors to avoid fitting to the noise, or creating un-realistic source models that are equally consistent with the data. The first prior on the source structure is through a regularization term, of strength $\bmath{\lambda_{\rm s}}$ and form $\bmath{\mathsf{R}}$, which ensures that the pixel-to-pixel variations of the source are smooth, but allows more structured sources when motivated by the data. The second prior comes from the lens model parameters, which results in a correlation of the image-plane surface brightness via the lens equation; given that gravitational lensing conserves the surface brightness, the image-plane emission cannot be random, but must result in regions of surface brightness that map directly back to the same position in the source plane.

The best source $\bmath{s}$ and lens model parameters $\bmath{\eta}$ are determined by maximising the posterior probability,
\begin{equation}
    P(\bmath{s}\, |\, \bmath{d},\, \bmath{\eta},\, \bmath{\lambda_{\rm s}},\, \bmath{\mathsf{R}}) = \frac{ P(\bmath{d}\, |\, \bmath{s},\, \bmath{\eta})\, P (\bmath{s}\, |\, \bmath{\lambda_{\rm s}},\, \bmath{\mathsf{R}}) }{ P(\bmath{d}\, |\, \bmath{\lambda_{\rm s}},\, \bmath{\eta},\, \bmath{\mathsf{R}})},
    \label{eq:bayes}
\end{equation}
where $P (\bmath{s}\, |\, \bmath{\lambda_{\rm s}},\, \bmath{\mathsf{R}})$ represents the prior on the source surface brightness distribution. The likelihood, in terms of the source and lens model parameters, is given by $P(\bmath{d}\, |\, \bmath{s},\, \bmath{\eta})$, and is assumed to be Gaussian, such that,
\begin{equation}
    P(\bmath{d}\, |\, \bmath{s},\, \bmath{\eta}) = \frac{1}{Z_D} \exp \left( -\frac{\chi^2}{2} \right).
    \label{eq:likelihood}
\end{equation}
The goodness of fit is determined via,
\begin{equation}
    \chi^2 = (\bmath{\mathsf{D}}\, \bmath{\mathsf{L}}(\bmath{\eta})\, \bmath{s} - \bmath{d})^T\,\bmath{\mathsf{C}}^{-1}\,(\bmath{\mathsf{D}}\, \bmath{\mathsf{L}}(\bmath{\eta})\, \bmath{s} - \bmath{d}),
    \label{eq:chi}
\end{equation}
and the normalization is given by
\begin{equation}
    Z_D = \sqrt{\det(2\pi\,\bmath{\mathsf{C}})}. 
    \label{eq:norm}
\end{equation}
Here, the psf and the Fourier transform of the data visibilities are used to evaluate the $\chi^2$ without explicitly entering into the visibility space.

The parametrization of the lens modelling is clearly important for determining the image- and source-plane surface brightness distribution, which is the focus of a companion paper that uses these data to investigate the mass distribution of the lens (Powell et al., in prep). To summarize, the mass model is an ellipsoidal power-law  with shear, and includes additional mass structure as described by a multipole expansion to explain the large-scale complexity (e.g. \citealt{powell2022,stacey2024}), and two localised mass concentrations to explain small-scale perturbations (e.g. \citealt{vegetti2012}) in the mass distribution. Finally, pixelated potential corrections are used to provide a well-focused source \citep{vegetti2009}.

\section{Results}
\label{sec:results} 

In this section, we present our results in the form of the deconvolved imaging of the data and an analysis of the resolved surface brightness distribution of the lensed radio emission, after the effect of gravitational lensing has been corrected for.

\subsection{Image-plane deconvolution}

In Fig.~\ref{fig:1938model}, we present the deconvolved image for the lens system. This has been made by first subtracting the model from the data visibilities, as determined by taking the Fourier transform of the image-plane model surface brightness distribution. We then take the Fourier transform of the residual visibilities to form the residual image, and then add the model surface brightness distribution convolved with the restoring beam. This is essentially equivalent to the process used to make {\sc clean} images, except that we have determined the sky model using the Bayesian forward modelling deconvolution process described above.
 
We see the radio emission from JVAS~B1938+666 that was previously seen at lower angular resolution \citep{king1998,spingola2020} has been well-recovered on VLBI-scales. In particular, the surface brightness distribution is dominated by an extremely thin gravitational arc that connects three of the lensed images from the quadruply-imaged part of the source. Also, the two merging lensed images to the west are recovered with a high signal-to-noise ratio, with a bright gravitational arc that extends around 200 mas. The two lensed images that make up the doubly-imaged part of the source are also well recovered and show compact structure with a fainter extended component to the south. The new VLBI data presented here have significantly better image fidelity, when compared to the previous VLBI imaging of \citet{king1997}, which did not recover the extended gravitational arc and produced a rather disjointed emission region for the two merging lensed images to the west. Never before has such a well-defined gravitational arc been detected on mas-scales (c.f. \citealt{more2009} and \citealt{spingola2019} for the case of MG J2016+112, and \citealt{spingola2018} for the case of MG~J0751+2716). This is due to the excellent sensitivity and $uv$-coverage provided by modern-day VLBI arrays. 

Based on the model surface brightness distribution, we measure a flux density at 1.7 GHz of $S_{\rm 1.7~GHz} = 490\pm49$~mJy for JVAS~B1938+666 (including a conservative 10 per cent error for the absolute flux density calibration of the global VLBI array). This is in good agreement with a previous measurement at 1.612 GHz \citep{king1998}.

\subsection{Source-plane morphology and properties}

\begin{figure}
    \centering
    \includegraphics[height=0.29\textheight]{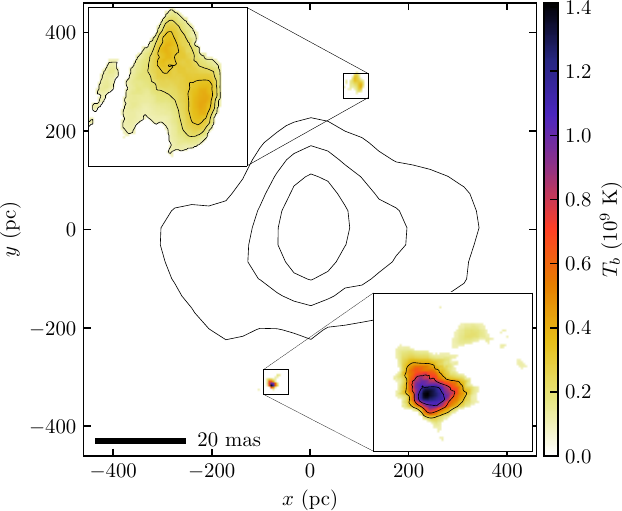}
    \caption{The source-plane brightness temperature distribution of JVAS B1938+666. The surface brightness of the rest-frame 0.7~$\mu$m emission from the host galaxy is shown with the black contours at $(0.25, 0.5, 0.75) \times I_{\rm 0.7~\mu m}$, the peak optical brightness. The two insets are $50\times50$~pc$^2$ in area and provide a zoom-in on the two radio components. The inset contours are $(0.25, 0.5, 0.75) \times T_{b,{\rm peak}}$, the peak brightness temperature of each radio component. We apply a mask to noise features with $T_b < 10^8$~K. The centre of the image is ${\rm RA}=19^{\rm h}38^{\rm m}25\fs3229$, ${\rm Dec}= +66\degr48\arcmin52\farcs871\,36$ (J2000).}
    \label{fig:1938src}
\end{figure}

\begin{figure}
    \centering
    \includegraphics[height=0.35\textheight]{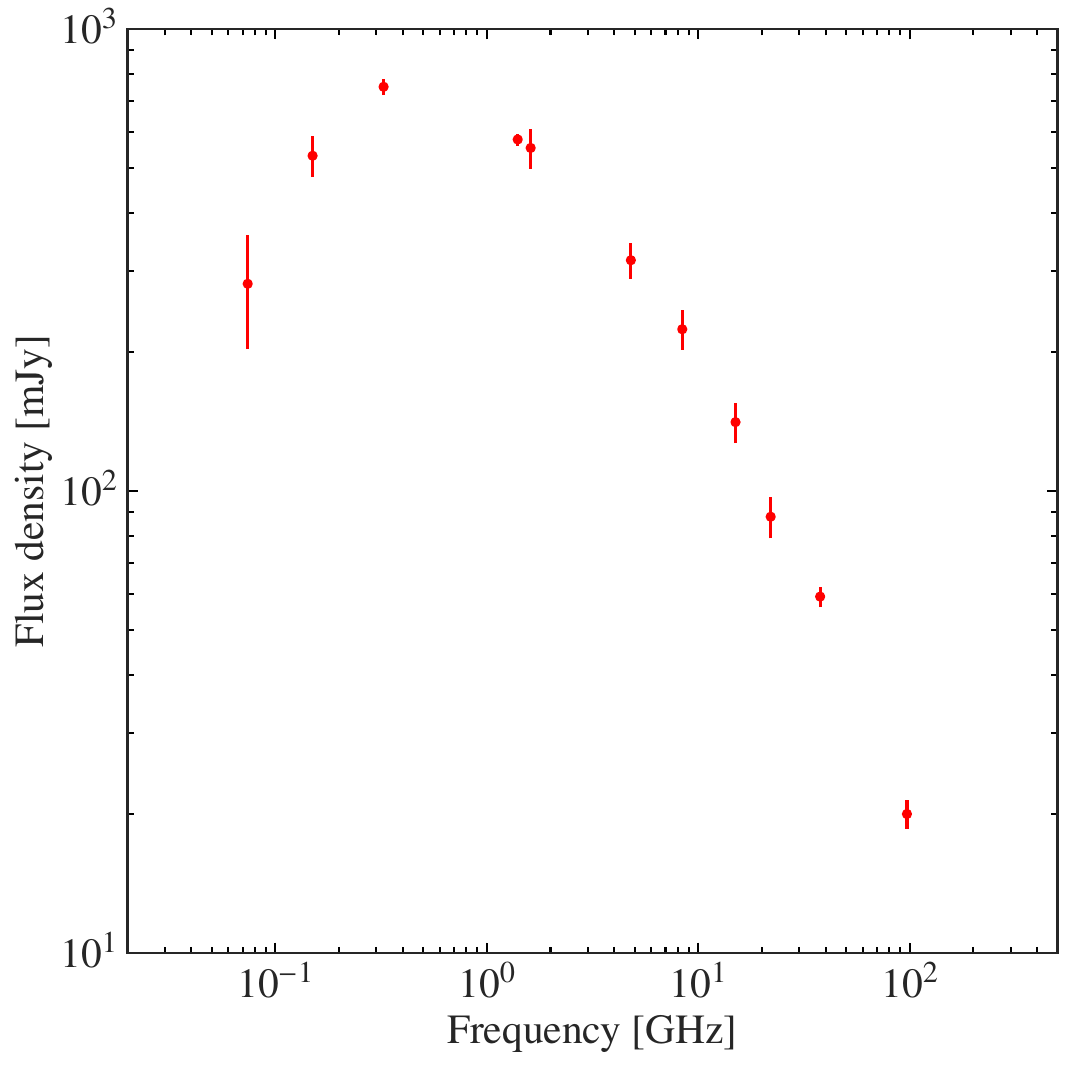}
    \caption{The radio spectral energy distribution of JVAS B1938+666 between 74 MHz and 97 GHz, which shows a clear turnover towards low frequencies. The data are taken from \citet{stacey2018}, and references therein. In addition, we include $S_{\rm 74~MHz} = 281\pm72$~mJy (VLA Low-Frequency Sky Survey; \citealt{cohen2007}), $S_{\rm 150~MHz} = 531\pm54$~mJy (Tata Institute of Fundamental Physics Giant Metrewave Radio Telescope Sky Survey Alternative Data Release; \citealt{intema2017}) and $S_{\rm 325~MHz} = 750\pm31$~mJy (Westerbork Northern Sky Survey; \citealt{rengelink1997}).
    } 
    \label{fig:1938spec}
\end{figure}

In Fig.~\ref{fig:1938src}, we present the surface brightness distribution of the lensed radio source, with the emission from the host galaxy at 2.1~$\mu$m (observed-frame; \citealt{spingola2020}) shown for reference. Also, to aid in the physical interpretation, we represent the intensity as a brightness temperature, $T_b$, which is determined using,
\begin{equation}
    T_b = \left( \frac{c^2}{2k} \right) \left( \frac{S_\nu}{\nu^2 \Omega} \right) (1+z),
\end{equation}
where $c$ is the speed of light, $k$ is the Boltzmann constant, $S_\nu$ is the flux density at frequency $\nu$, $\Omega$ is the pixel solid angle and $z$ is the redshift. Note that this is a slightly modified version of the commonly used expression for the brightness temperature because we have a pixelated model, as opposed to one that is convolved with an elliptical Gaussian function \citep{condon1982}. In addition, we have calculated the total rest-frame luminosity density at 1.7 GHz using,
\begin{equation}
    L_{\rm 1.7~GHz} = 4\pi D_L^2 S_{\rm 1.7~GHz} (1+z)^{-(1+\alpha)},
\end{equation}
where $D_L$ is the luminosity distance. Here, the spectral index $\alpha$ is defined by the power-law,
\begin{equation}
    S_\nu = S_0 \left( \frac{\nu}{\nu_0} \right)^{\alpha}.
    \label{eq:spec}
\end{equation}
In Fig.~\ref{fig:1938spec}, we present the radio spectral energy distribution of JVAS B1938+666, which shows a turnover with a clearly varying spectral index. Therefore, we fit equation (\ref{eq:spec}) to three different parts of the radio spectrum, finding, $\alpha_{0.074}^{0.325} = +0.59\pm0.13$, $\alpha_{1.4}^{4.8} = -0.49\pm0.03$ and $\alpha_{4.8}^{37.7} = -0.79\pm0.05$, which correspond to the low-, mid- and high-frequency spectral index, respectively (the upper and lower indices of $\alpha$ represent the frequency bounds in GHz used for fitting).

We see from Fig.~\ref{fig:1938src} that the radio morphology of JVAS~B1938+666 shows two distinct components separated by 642~pc in an almost north-south direction, with the rest-frame optical emission located centrally in between; this source-plane morphology is consistent with the previous parametric lens modelling of the radio emission by \citet{king1998} and \citet{spingola2020}. Also, a previously imaged CO (1--0) molecular gas disc lies between the two radio components and is coincident with the rest-frame optical emission \citep{spingola2020}, which we assume is the centre of the gravitational potential of the system. The radio component to the south results in the quadruply-imaged emission and the extended gravitational arc seen in Fig.~\ref{fig:1938model}. This component has the highest intensity of the radio source, with a peak brightness temperature of $10^{9.15}$~K and a diffuse region with a brightness temperature that extends smoothly down to $10^{8.45}$~K. The southern component has a projected area (as defined within a factor 0.25 of the peak brightness) of 231~pc$^{2}$ and a maximum extent of about 20 pc. The radio component to the north, which results in the doubly-imaged emission seen in Fig.~\ref{fig:1938model}, has a more extended morphology with two compact sub-components that are separated by 20-pc in projection, and a faint extension to the south-east. The projected area of the northern radio component is 886~pc$^2$. The two compact sub-components have a similar brightness temperature of around $10^{8.60}$~K, whereas the fainter extension has a brightness temperature of $10^{8.0}$~K. The total flux density of the model surface brightness distribution is $11.9\pm1.2$~mJy, which corresponds to a luminosity density of $10^{26.34\pm0.04}$~W~Hz$^{-1}$, when using the mid-frequency spectral index.

\section{Discussion}
\label{sec:disc}

We now characterise the background radio source, based on the resolved morphology and properties that we have recovered here using Bayesian forward modelling deconvolution, and from the previous multi-wavelength observations of JVAS~B1938+666.

\subsection{Radio source classification}

First, it is important to note that the two radio sources we observe could be part of a multi-plane lens system, where the infra-red emitting source located in between could be a second lensing galaxy at a higher redshift. This has recently been shown to be the case for the lens system PS J1721+8842 \citep{dux2024}, which also has six lensed images and was originally thought to be a dual AGN system that was lensed by a single-plane lensing galaxy \citep{mangat2021,lemon2022}. However, from our resolved imaging, we see that the morphology and surface brightness of the two radio sources are not the same, and so, they must be two distinct regions of radio emission. Given that the observed radio, infra-red and CO (1--0) molecular gas image-plane surface brightnesses can be explained by the same lens model, we confirm that all components are at the same redshift of 2.059. Second, the high brightness temperatures of the radio emitting regions cannot be consistent with star-formation; a radio source with the mid-frequency spectral index and redshift of JVAS~B1938+666 has a maximum theoretical brightness temperature of $T_b\leq10^{5.4}$~K from an optically thick star-forming region \citep{condon1991}. Therefore, we conclude that the radio emission is associated with non-thermal AGN activity.

Given the edge-brightened morphology of the northern and southern radio components (bright compact emission with weaker extended emission towards the centre) with respect to the AGN host galaxy, we interpret these two features as mini-lobes containing hotspots. This is consistent with the lack of variability seen from a 3-month monitoring of JVAS~B1938+666 with the Very Large Array at 8.46 GHz \citep{rumbaugh2015} and two observations at 37.7 GHz separated by about 1~yr \citep{spingola2020}; we note that if the radio source were a core-jet object, where the southern component is the likely core given its flatter radio spectrum ($\alpha_{1.612}^5 = -0.5$; \citealt{king1997}) and higher brightness temperature, then we would be more likely to see variability. Given the small angular size of 642~pc, edge-brightened emission with hotspots, peaked radio spectrum with a relatively steep high-frequency radio spectral index, and the lack of any variability, we classify this radio source as a compact symmetric object (CSO; \citealt{wilkinson1994,kiehlmann2024a}) of type 2 \citep{readhead2024}. However, we note that the component separation is slightly larger than the current upper bound for this class of CSO, which is around 500~pc.

A potential caveat to this classification is the lack of evidence for a radio core at the centre of the host galaxy in our VLBI imaging or in any other deep imaging of this system made to date. However, the radio cores of CSOs are typically very weak and not often detected \citep{readhead1996}. We place a conservative 10-$\sigma$ limit of $<0.3$~mJy for a point-like radio core in the image plane, which corresponds to $<0.2$~per cent of the total radio emission being associated with a core in the source plane. Alternatively, the component to the south could be a radio core, where that part of the optical host galaxy is heavily enshrouded by dust. Such offsets between the dust and the rest-frame optical emission have been seen in mm-bright starburst galaxies \citep{rybak2015a}, but the CO molecular gas from low to high excitation is co-spatial with the dust in those cases \citep{rybak2015b}. Although JVAS~B1938+666 shows significant emission from heated dust \citep{stacey2018}, an obscured core-jet scenario is unlikely, given the location of the CO (1--0) molecular gas in this system. Therefore, the CSO classification is likely robust. This could be confirmed by showing that the jet-expansion speed is less than 2.5c, since this implies a viewing angle of more than 45 deg and a Lorentz factor that is less than 3 \citep{kiehlmann2024a}. This will be tested using a second epoch of VLBI imaging to determine whether either of the two radio components exhibits superluminal motion, or not, over a 15 yr period.  

\subsection{Comparison with a bow shock model for CSOs}

CSOs are generally considered to be short-lived radio sources with ages of around $10^{4-5}$ yr, given their small projected angular sizes and slow expansion speeds \citep{polatidis2003}. Should they continue to expand then they are expected to evolve into FRII type radio sources \citep{readhead1996}. Therefore, studying CSOs at high redshift can provide some insight to the early-phases of AGN activity when their host galaxies are still forming. The host galaxy of JVAS B1938+666 is rather compact (effective radius of $R_e = 460 \pm 70$~pc), is red in colour, and is dusty ($L_{8-1000~{\rm \mu m}} = 10^{12.1}$~L$_{\odot}$) with a large amount of cold molecular gas ($3.4 \pm 0.8 \times 10^{10}$~M$_\odot$) in a 1.5 kpc-scale disc \citep{stacey2018,spingola2020}. 

These observed properties are in good agreement with the model proposed by \citet*{bicknell1997}, who predicted that the slow expansion speeds and short-lived nature is due to the ISM of the dusty host galaxies being rather dense, with cold molecular gas masses of 10$^{9-11}$~M$_{\odot}$. In this model, the jets expand into the ISM to form a bow shock that fragments the molecular clouds. This fragmentation can lead to bursts of star-formation, and indeed, the host galaxy of JVAS~B1938+666 is thought to be forming stars at a rate of about 700~M$_{\odot}$~yr$^{-1}$ \citep{stacey2018}. Also, the radio jet can produce a transverse shock, perpendicular to the flow direction of the jet, which may also explain the increased brightness of the CO (1--0) emission seen to the western part of the gas disc \citep{spingola2020}. One potentially interesting feature of our VLBI imaging is the detection of two hotspots in the northern radio component. Given that hotspots show the termination point of the jet, this suggests that the ISM is sufficiently dense to first produce a termination point, but also that the jet has changed direction. This "dentist drill" model for the jet is also a prediction by \citet{bicknell1997} for a dense ISM. The resulting bow-shock from the jets is also expected to collisionally ionize the ISM, which can result in depolarization and Faraday rotation. The southern radio component of JVAS B1938+666 shows significant depolarisation from 15 and 5 GHz to 1.612 GHz, where the polarisation fraction changes from around 15 to 1.5 per cent, with a rotation measure of around 500 rad~m$^{-2}$ \citep{king1997}. This provides strong evidence for an ionised medium, in agreement with the bow-shock model.

Whether JVAS B1938+666 evolved into an FRII-type radio source is not clear. A recent study of a well-selected sample of CSOs has suggested that such radio sources are triggered via the interaction of a star with a black hole, and that their small angular-size is due to their being insufficient fuel to further drive the central engine \citep{readhead2024}. The morphology of JVAS~B1938+666 is of a type 2 CSO, where mini-lobes and hotspots have formed, which are not expected to expand further. Therefore, further study of the ISM density and pressure around this system may help determine whether the small angular size of this radio source is due to the local environment containing the AGN or the lack of available fuel to maintain the jet expansion.

\section{Conclusions}
\label{sec:conc}

Gravitational lensing has long promised to provide a high angular resolution view of the Universe, which was often limited by our ability to model the deflection of the light to a high enough precision. The advancements in constructing complex mass models for the lens, via both parametric \citep{stacey2024} and non-parametric \citep{vegetti2009} components, has been driven by the development of sophisticated modelling algorithms (e.g. \citealt{powell2021}) and high quality data sets \citep{spingola2018} that test the lens structure on mas-scales (equivalent to around 6 to 8 pc in the lens-plane, for a lens redshift between 0.5 and 1). Here, we have applied a Bayesian forward modelling procedure to determine the complex surface brightness distribution in both the image and source plane of a lensed CSO at redshift 2.059. The high quality of the global VLBI imaging presented here has provided the necessary constraints to derive a precise mass model for the lens system (presented in a companion paper), from which the radio source could be well-resolved. We find that the morphology and multi-wavelength properties of the CSO and its host galaxy are consistent with the bow-shock model for jetted compact radio sources \citep{bicknell1997}.

It is important to note that gravitational lensing provides a biased view of the high redshift Universe due to differential magnification, and so, the lensing configuration plays an important role in determining a robust surface brightness distribution for the reconstructed source. It is for this reason that lens systems with Einstein rings and extended gravitational arcs provide the best targets for an analysis such as the one presented here (see also \citealt{stacey2025,deroo2025} for recent applications of this method to data from the Atacama Large Millimetre Array). Also, the Bayesian forward modelling technique is best suited to highly resolved and structured sources that dominate the visibility function from the interferometer. Although future instruments, like the Square Kilometre Array (SKA), will find around $10^5$ lensed radio sources \citep{mckean2015}, it will only be through combining the SKA within a VLBI network that the high redshift Universe will be properly resolved with gravitational lensing on the angular-scales needed to test various models for galaxy formation and evolution.

\section*{Acknowledgements}

We thank Neal Jackson for their positive and constructive review of our paper. This work is based on the research supported in part by the National Research Foundation of South Africa (Grant Number: 128943). SV thanks the Max Planck Society for support through a Max Planck Lise Meitner Group. DMP and SV have received funding from the European Research Council (ERC) under the European Union?s Horizon 2020 research and innovation programme (grant agreement No 758853). CS acknowledges financial support from INAF under the project "Collaborative research on VLBI as an ultimate test to $\Lambda$CDM model" as part of the Ricerca Fondamentale 2022. This work was supported in part by the Italian Ministry of Foreign Affairs and International Cooperation, grant number  PGRZA23GR03. The European VLBI Network is a joint facility of independent European, African, Asian, and North American radio astronomy institutes. The scientific results from the data presented in this publication are derived from the following EVN project code(s): GM068. The National Radio Astronomy Observatory is a facility of the National Science Foundation operated under cooperative agreement by Associated Universities, Inc.

%%%%%%%%%%%%%%%%%%%% REFERENCES %%%%%%%%%%%%%%%%%%

\section*{Data Availability}

The data used here are publicly available via the EVN archive.

%%%%%%%%%%%%%%%%%%%% REFERENCES %%%%%%%%%%%%%%%%%%

% The best way to enter references is to use BibTeX:

\bibliographystyle{mnras}
\bibliography{references}

%%%%%%%%%%%%%%%%%%%%%%%%%%%%%%%%%%%%%%%%%%%%%%%%%%

%%%%%%%%%%%%%%%%% APPENDICES %%%%%%%%%%%%%%%%%%%%%

%\appendix

%\section{Some extra material}

%If you want to present additional material which would interrupt the flow of the main paper, it can be placed in an Appendix which appears after the list of references.

%%%%%%%%%%%%%%%%%%%%%%%%%%%%%%%%%%%%%%%%%%%%%%%%%%

% Don't change these lines
\bsp	% typesetting comment
\label{lastpage}
\end{document}